# ON THE MAGNETICALLY DRIVEN FERROELECTRIC PHASE IN GdMnO$_3$

J.L. RIBEIRO

Centro de Física, Universidade do Minho, Campus de Gualtar, 4710-007 Braga, Portugal

*At room temperature, GdMnO3 is a paraelectric and paramagnetic with a distorted perovskite structure of orthorhombic symmetry (space group Pnma). On cooling, it undergoes a phase transition sequence to a magnetic incommensurate phase ($\vec{k} = \delta(T)\vec{a}^*$; $T_{c1}$=42K) and a A-type antiferromagnetic phase ($T_{c2}$=27K). At low temperatures ($T \leq 12K$), a magnetic field applied along the a-axis destabilizes the antiferromagnetic phase and induces a first order transition to a magnetic commensurate modulated phase ($\vec{k} = \frac{1}{4}\vec{a}^*$) that is also ferroelectric ($\vec{P} // \vec{c}$). This work analyses this field induced phase transition from the point of view of the symmetry and Landau theory.*

## Introduction

The search for novel single phase materials displaying an enhanced magneto-electric effect attracts increasing interest, not only because of the potential technological application that can be envisaged, but also for the subtle physical mechanisms that are involved [1,2]. Among this class of materials, the rare-earth manganites RMnO3 (R=Gd, Tb and Dy) are perhaps the simpler prototypes that show an unusual interplay of magnetic and electric properties [3-7].

At room temperature, these materials show a reference phase with a distorted orthorhombic perovskite structure (space group Pnma). Among the three compounds, GdMnO3 is the one that shows the less distorted structure and the simpler phase diagram. At about $T_{c1}$=42K, it undergoes a phase transition from the reference phase to a magnetic incommensurate phase with a modulation wavevector directed along the $\vec{a}-axis$ ($\vec{k}=[\delta,0,0]$; $\delta(T_{c1}) \approx 0.24$) [4,8]. In this phase, the magnetic modulation corresponds to a collinear arrangement of the Mn spins along the $\vec{a}-axis$, ferromagnetically coupled within the ac-plane and antiferromagnetically aligned along the $\vec{b}-axis$. On cooling below $T_{c1}$, the modulation wavenumber decreases and approaches the commensurate value $\delta = 0.2$ [8]. However, this commensurate value is not reached because the system undergoes a discontinuous phase transition to a A-type antiferromagnetic phase at about Tc2=23K. This low temperature phase corresponds to a lock-in at the point $\delta = 0$ [8].

Along this phase sequence, at zero magnetic field, GdMnO3 remains paraelectric. However, the application of a strong magnetic field below $T^*$~12K along the $\vec{a}-axis$ induces a ferroelectric phase with a polarization oriented along the $\vec{c}-axis$. Structural studies [8] indicate that this magnetically induced ferroelectric phase is accompanied by a commensurate modulation of the Mn momenta corresponding to a wavevector $\vec{k}=\left[\frac{1}{4};0;0\right]$. A similar situation is also found in TbMnO$_3$ and DyMnO$_3$. Here, even if the zero field phase diagram is more complex, a commensurate plateau with $\vec{k}=\left[\frac{1}{4};0;0\right]$ and a ferroelectric polarization directed along the c-axis are again stabilized by a magnetic field applied along the a-axis [3,4,7].

The purpose of the present work is to analyse the field induced stabilization of the ferroelectric phase in GdMnO3 from the point of view of the symmetry. This analysis will be based on general symmetry considerations and aimed at clarifying why the polarization is related to a particular value of the modulation wavenumber. In addition, the method adopted, will allow us to derive a simple set of symmetry based free energy functionals capable of describing the phenomenology observed in GdMnO$_3$

**Can an irreducible magnetic order parameter induce ferroelectricity?**

The magnetic modulation observed in GdMnO3 below $T_{c1}$ corresponds to a simple collinear modulation of the magnetic momenta of the Mn ions that likely results from the condensation of an irreducible magnetic order parameter. Let us then consider the case of a irreducible modulated order parameter that is characterized by a single wavevector $\vec{k}$ located in the interior of the Brillouin zone and directed along a line of fixed symmetry (which corresponds to the $\Sigma$ line, in the case of the RMnO$_3$ compounds). This modulated order parameter can be written as:

$$\vec{S}(\vec{x}) = \vec{S}e^{i\vec{k}\cdot\vec{x}} + \vec{S}^*e^{-i\vec{k}\cdot\vec{x}} \qquad (1)$$

where, $\vec{S} = \vec{S}_0 e^{i\pi\Phi}$ and $\Phi$ is a phase, defined with respect to the underlying discrete lattice. Let us assume that $\vec{S}(\vec{x})$ is of a magnetic nature and, therefore, is odd under the operation of time reversal $\theta$, (that is, $\theta\vec{S}(\vec{x}) = -\vec{S}(\vec{x})$). Under which conditions can this general modulated order parameter give rise to a spontaneous polarization $P$?

This problem is similar to the one raised by the improper ferroelectricity induced by displacive lattice modulation and, with the necessary adaptation, can be tackled along the same lines. In terms of Landau theory, for example, one can ask if $\vec{S}(\vec{x})$ allows the construction of mixed invariants linear in the electric polarization $P$, a necessary condition for improper ferroelectricity [9]. It can be easily seen that this possibility can never be realized when the modulation is incommensurate [9,10]. In the case of a commensurate modulation, however, the answer to the question requires the analysis of the transformation properties of the translational invariants that can be constructed from the components of the order parameter (note that $P$ is itself a translational invariant). If, under the symmetry operations of the reference phase, any of these possible invariants is transformed as a polar vector, then a mixed invariant linear in $P$ can be constructed. Consequently, improper ferroelectricity will be allowed by symmetry.

This method of analysis requires the knowledge of the way the different elements of the symmetry group of the reference phase act on the linear space generated by the components of the order parameter. In the case of an irreducible order parameter, this amounts to knowing the complete irreducible co-representations (CICR) of the paramagnetic space group and the standard methods can be used to construct them [11-14].

For the case pertaining to GdMnO$_3$ (paramagnetic space group (Pnma)´ and modulation wavevector $\vec{k} = (\delta(T),0,0)$, k$_7$ in Kovalev`s tables), the CICR matrices can readily be obtained. These are listed in Table 1 for the generators of the magnetic space group ($C_{2x}$, $\sigma_y$, $i$ and $i\theta$). The matrices for the other elements of the magnetic space group can be obtained using the multiplication table of the group and taking into account that the anti-unitary operations conjugate the coefficients of the matrices upon which they act.

The method outlined also requires a systematic way to construct the possible translational invariants from the components of the order parameter $\vec{S}(\vec{x})$. In the present case, given that $S$ and $S^*$ are complex numbers, the analytical form of these translational invariants is simply determined by the image that the translational sub-group of the reference phase induces in the complex plane. For a commensurate wavenumber $\delta = \dfrac{n}{m}$, this image is isomorphic to the group *Cm* [10]. Therefore, as shown in [9-10], any homogeneous polynomial defined in the space of the complex numbers that is invariant under *Cm* (requirement of translational invariance) must necessarily be expressed as linear combinations of terms of the type $S_0^m \cos(m\Phi)$ and $S_0^m \sin(m\Phi)$. This means that one can determine if mixed terms linear in P are allowed simply by checking how these translational invariants are transformed under the remaining symmetry operations of (Pnma)´.

Because of the different parity under spatial inversion of the two types of translational invariants, a coupling term linear in P must necessarily be of the form $PS_0^m \sin(m\Phi)$. Moreover, because $\theta P = P$ and $\theta S_0 = -S_0$, such a mixed term will be invariant under time reversal if only if *m* is even. That is, for the particular symmetry under analysis, only magnetic commensurate phases of the type $\delta = \frac{odd}{even}$ can support a ferroelectric polarisation.

Conversely, the translational invariants of the form $S_0^m \cos(m\Phi)$, which are even under spatial inversion and odd under time reversal if *m* is odd, will allow for mixed invariants linear in a magnetization *M*. In this case, an homogeneous magnetization may co-exist with the modulated magnetic order.

Table 2 shows the cases in which the translational invariants of the magnetic order parameter $\vec{S}(\vec{x})$ transform as components of a magnetization or a polarization. As seen, in this latter case, a polarization directed along the $\vec{c} - axis$ is the unique possibility. Independently of the symmetry of the primary order parameter, it is allowed only if $\delta = \frac{odd}{even}$.

## III. The field induced ferroelectricity in GdMnO3

As seen, at low temperatures and zero field, the ground state of GdMnO3 corresponds to a A-type antiferromagnetic order in which the Mn spins are ferromagnetically ordered in the $ac$-plane (directed along the $\vec{a}-axis$) and antiferromagnetically aligned along the $\vec{b}-axis$. If we denote by $M_{a,b}$ the magnetization of the two sub-lattices, we can write the free energy density as:

$$F = f(M_a^2 + M_b^2) + gM_aM_b + 2h(M_a^4 + M_b^4) - (M_a + M_b)H \qquad (2)$$

A coefficient $g > 0$ will favour the observed antiferromagnetic coupling between neighbour $ac$-planes. In terms of the ferro- ($M$) and antiferromagnetic ($\xi$) order parameters, defined as $M = M_a + M_b$ and $\xi = M_a - M_b$, this free energy can be expressed as:

$$F = \frac{1}{2}(f + \frac{g}{2})M^2 + \frac{1}{2}(f - \frac{g}{2})\xi^2 + \frac{h}{4}(M^4 + \xi^4 + 6M^2\xi^2) - MH \qquad (2')$$

It is well known that the antiferromagnetic order can be destabilised by an external magnetic field. For example, by imposing to (2´) the equilibrium conditions $\frac{\partial F}{\partial M} = \frac{\partial F}{\partial \xi} = 0$, one obtains:

$$M = \frac{H}{(f + \frac{g}{2}) + 3h\xi^2 + hM} \qquad (3a)$$

$$\xi^2 = \frac{\frac{g}{2} - f}{h} - 3M^2 \qquad (3b)$$

As seen, the magnetization induced by the external field reduces the amplitude of the antiferromagnetic order parameter $\xi$. Eventually, a high enough field ($H \geq H_{c1}$) can give rise to a transition to a ferromagnetic phase ($\xi=0$; $M \neq 0$).

However, in GdMnO3 there is a competition between ferro- and antiferromagnetic interactions within the *ac*-planes. This competition, seen for example by the stabilization of the incommensurate phase in the range $T_{c2}<T<T_{c1}$, may favour the onset of a modulated phase under field. That is, above a certain threshold $H_{c2} < H_{c1}$ and alternatively to the field induced rotation of the whole magnetization of a sub-lattice, the system may prefer to recover a spin modulation within each *ac*-plane. In such a case, a key point would be to understand why the commensurate modulation $\delta = \frac{1}{4}$ is energetically favourable (and consequently a polarization along the $\vec{c}-axis$ stabilised).

As seen in the previous section, for $\delta = \frac{odd}{odd}$ or $\delta = \frac{even}{odd}$ (that is, when *m* is odd), the free energy density corresponding to commensurate phase includes a nomial that is linear in a magnetization *M*. Let us consider a system, like GdMnO3, with two magnetic sub-lattices and denote by $S_{a,b}$ and $M_{a,b}$ the amplitude of the commensurate spin modulation and the homogeneous magnetization in the sub-lattices *a* and *b*, respectively. Then, the free energy of the system can be written as:

$$F_2 = \frac{\alpha}{4}(S_a^2 + S_b^2) + \frac{\beta}{4}(S_a^2 + S_b^2) + \nu\cos(m\Phi)[M_a S_a^m + M_b S_b^m] + \frac{\gamma}{2}(S_a^{2m} + S_b^{2m})\cos(2m\Phi) + $$
$$+ \Omega(S_a^2 M_a^2 + S_b^2 M_b^2) + \frac{1}{2\chi_m}(M_a^2 + M_b^2) + g_1 M_a M_b + h(M_a^4 + M_b^4) - (M_a^x + M_b^x)H_x$$

(4)

Here, the contribution of the normal Umklapp term $[\frac{\gamma}{2}(S_a^{2m} + S_b^{2m})\cos(2m\Phi)]$ along with other trivial invariants are included. Note that the direction of $M_{a,b}$ is dictated by the symmetry of the order parameter and by the type of the wavenumber $\delta$ (see Table 2). It is also assumed that the field is applied along the *a-axis* and $g_1>0$.

Naturally, the mixed term $\nu\cos(m\Phi)[M_a S_a^m + M_b S_b^m]$ in equ. 4 is the lowest order term that may favour the particular commensurate modulation. Because $m$ is odd and the antiferromagnetic coupling $S_a = -S_b = S$ preferred, such term can be can be written as:

$$\nu\cos(m\Phi)[M_a S_a^m + M_b S_b^m] = \nu S^m \cos(m\Phi)[M_a - M_b] = \nu S^m \cos(m\Phi)\xi \quad (5)$$

That is, the modulated order parameter $S$ is necessarily coupled to the antiferromagnetic order parameter $\xi$. Because of this reason, this type of commensurate phase will be destabilised by the external field. In fact, like in the case of the homogeneous antiferromagnetic order, the amplitude of $\xi$ will decrease under an external field as:

$$\xi^2 \approx \frac{g_1 - (2\Omega S^2 + \frac{1}{\chi_m})}{h} - 3\frac{H_x^2}{\left[\frac{3}{2}g_1 - (2\Omega S^2 + \frac{1}{\chi_m})\right]^2} \quad (6)$$

Consequently, the lowest order lock-in term that can eventually favour the modulated phase will decrease (proportionally to $\xi$) as the external field increases and to relative stability of the phase diminish.

This effect of the external field is necessarily absent if $\delta = \frac{odd}{even}$ because, in this case, the symmetry requires that $M_a = M_b = 0$. Here, the potential secondary order parameter is an electric polarization $P$ (oriented along $\vec{c}$, as seen) and the free energy density for a system with two sub-lattices is:

$$F_3 = \frac{\alpha'}{4}(S_a^2 + S_b^2) + \frac{\beta'}{4}(S_a^2 + S_b^2)^2 + \eta \sin(m\Phi)\left[P_a S_a^m + P_b S_b^m\right] + \frac{\gamma}{2}(S_a^{2m} + S_b^{2m})\cos(2m\Phi) +$$
$$+ \Omega_P(S_a^2 P_a^2 + S_b^2 P_b^2) + \frac{1}{2\chi_P}(P_a^2 + P_b^2) + g_2 P_a P_b + h_2(P_a^4 + P_b^4) - (M_a^x + M_b^x)H_x + \quad (7)$$
$$\Omega_m(S_a^2 M_a^2 + S_b^2 M_b^2) + \frac{1}{2\chi_m}(M_a^2 + M_b^2)$$

Note that in this case ($m$=even), the lowest order lock-in term can be expressed as $\eta \sin(m\Phi)\left[P_a S_a^m + P_b S_b^m\right] = \eta S^m \sin(m\Phi)\left[P_a + P_b\right] = \eta S^m \sin(m\Phi)P$. That is, the primary order parameter is linearly coupled to $P=P_a+P_b$ (ferroelectric order parameter). Moreover, the intensity of this coupling is not affected by an external magnetic field and, if $\eta \sin(m\Phi) < 0$, favours the stability of the phase. Evidently, among all the possible values of the type $\delta = \frac{odd}{even}$ the modulation $\delta = \frac{1}{4}$ is the one that corresponds to the lowest possible lock-in order and is therefore expected to be more stable.

From equation 7, the electric polarization can be explicitly related to the amplitude of the magnetic modulation as:

$$P = \frac{-\eta S^m \sin(m\Phi)}{\left(\Omega_P S^2 + \frac{1}{2\chi_P} + \frac{g}{2}\right)} \quad (8)$$

In the above discussion the case of the stabilisation of an incommensurate modulation by the external field was not considered. In fact, this possibility corresponds to a case in which the relevant mixed lock-in invariants linear in P or M are forbidden by symmetry. On another hand, the phase transition induced by the field in the case of GdMnO3 occurs at temperatures well below the transition from the reference to the incommensurate phase. Thus, the amplitude of the potential order parameter is likely to be high enough to favour, through the interaction with the discrete lattice, commensurate modulations.

**REFERENCES**

[1]     Fiebig M: The revival of the magnetoelectric effect.  J. Phys. D: Appl. Phys. 2005; 38: R123- R152.

[2]     Eerenstein W, Mathur ND and Scott JF: Multiferroic and magnetoelectric materials. Nature  2006; 442: 759-765.

[3]     Kimura T, Goto T, Shintani H, Ishizaka K, Arima T and Tokura Y: Magnetic control of ferroelectric polarization. Nature 2003; 426: 55- 58.

[4]     Goto T,  Kimura T, Lawes G,  Ramirez AP, and Tokura Y: Ferroelectricity and Giant Magnetocapacitance in Perovskite Rare-Earth Manganites. Phys. Rev. Lett. 2004;  92: 257201.

[5]     Kadomtseva AM, Popov YF, Vorobev GP, Kamilov KI,  Pyatakov AP,  Ivanov VY, Mukhin AA, and Balashov AM: Specificity of magnetoelectric effects in a new $GdMnO_3$ magnetic ferroelectric . JETP Lett. 2005; 81: 19-23.

[6]     Noda K, Nakamura S, Nagayama J, and Kuwahara H. Magnetic field and external-pressure effect on ferroelectricity in manganites: Comparison between $GdMnO_3$ and $TbMnO_3$. J. Appl. Phys. 2005; 97: 10C103.

[7]     Kimura T, Lawes G, Goto T, Tokura Y and Ramirez AP: Magnetoelectric phase diagrams of orthorhombic $RMnO_3$ (R=Gd, Tb, and Dy). Phys. Rev. B 2005; 71:  224425.

[8]     Arima T, Goto T,Yamasaki Y, Miyasaka S, Ishii K, Tsubota M, Inami T, Murakami Y, and Tokura Y: Magnetic-field-induced transition in the lattice modulation of colossal magnetoelectric $GdMnO_3$ and $TbMnO_3$ compounds. Phys. Rev. B 2005; **72**: 100102R.

[9]     Tolédano JC and Tolédano P: The Landau Theory of Phase Transitions. Singapore, World Scientific; 1987.

[10]    Michel L: Minima of Higgs-Landau polynomials. In: Regards sur la Physique Contemporaine. Paris; CNRS; 1980, 157.

[11]    Kovalev O. V: Representations of the Crystalographic Space Groups- Irreducible Representations, Induced Representations and Corepresentations; Gordon and Breach Science Publishers; 1993.

[12]    Bradley CJ and Davies BL: Magnetic Groups and Their Corepresentations. Rev. Mod. Phys. 1968; 40: 359-379

[13]    E. P. Wigner: Group Theory and its Applications to Quantum Mechanics of Atomic Spectra. Academic Press;1960.

[14]    Wigner EP: Normal form of antiunitary operators. J. of Math. Phys. 1960; 1: 409-413.

**TABLE 1:** Matrices representing the generators of the group (Pnma)´ in the four of its complete irreducible co-representations at $\vec{k} = \delta \vec{a}_1^*$.

|  | $C_{2x}$ | $\sigma_z$ | $i\theta$ | $i$ |
|---|---|---|---|---|
| $\Gamma(A_1)$ | $\begin{bmatrix} \varepsilon & 0 \\ 0 & \varepsilon^* \end{bmatrix}$ | $\begin{bmatrix} \varepsilon & 0 \\ 0 & \varepsilon^* \end{bmatrix}$ | $\begin{bmatrix} -1 & 0 \\ 0 & -1 \end{bmatrix}$ | $\begin{bmatrix} 0 & 1 \\ 1 & 0 \end{bmatrix}$ |
| $\Gamma(B_2)$ | $\begin{bmatrix} -\varepsilon & 0 \\ 0 & -\varepsilon^* \end{bmatrix}$ | $\begin{bmatrix} -\varepsilon & 0 \\ 0 & -\varepsilon^* \end{bmatrix}$ | $\begin{bmatrix} -1 & 0 \\ 0 & -1 \end{bmatrix}$ | $\begin{bmatrix} 0 & 1 \\ 1 & 0 \end{bmatrix}$ |
| $\Gamma(A_2)$ | $\begin{bmatrix} \varepsilon & 0 \\ 0 & \varepsilon^* \end{bmatrix}$ | $\begin{bmatrix} -\varepsilon & 0 \\ 0 & -\varepsilon^* \end{bmatrix}$ | $\begin{bmatrix} -1 & 0 \\ 0 & -1 \end{bmatrix}$ | $\begin{bmatrix} 0 & 1 \\ 1 & 0 \end{bmatrix}$ |
| $\Gamma(B_1)$ | $\begin{bmatrix} -\varepsilon & 0 \\ 0 & -\varepsilon^* \end{bmatrix}$ | $\begin{bmatrix} \varepsilon & 0 \\ 0 & \varepsilon^* \end{bmatrix}$ | $\begin{bmatrix} -1 & 0 \\ 0 & -1 \end{bmatrix}$ | $\begin{bmatrix} 0 & 1 \\ 1 & 0 \end{bmatrix}$ |

**TABLE 2: Transformation properties of the lowest order translational invariants of a modulated order parameter.**

|  |  | $\Gamma(A_1)$ | $\Gamma(B_2)$ | $\Gamma(A_2)$ | $\Gamma(B_1)$ |
|---|---|---|---|---|---|
| $S^m \cos(m\Phi)$ | $\delta = \dfrac{odd}{odd}$ | $M_y$ | — | $M_z$ | $M_x$ |
|  | $\delta = \dfrac{even}{odd}$ | — | $M_y$ | $M_x$ | $M_z$ |
|  | $\delta = \dfrac{odd}{even}$ | — | — | — | — |
|  |  | $\Gamma_1$ | $\Gamma_2$ | $\Gamma_3$ | $\Gamma_4$ |
| $S^m \sin(m\Phi)$ | $\delta = \dfrac{odd}{odd}$ | — | — | — | — |
|  | $\delta = \dfrac{even}{odd}$ | — | — | — | — |
|  | $\delta = \dfrac{odd}{even}$ | $P_z$ | $P_z$ | $P_z$ | $P_z$ |